# Grey Power Models Based on Optimization of Initial Condition and Model Parameters

Yunchol Jong

[a]Center of Natural Science, University of Sciences, Pyongyang, DPR Korea

E-mail: yuncholjong@yahoo.com

**Abstract.** We propose a novel approach to improve prediction accuracy of grey power models including GM(1,1) and grey Verhulst model through optimization of the initial condition and model parameters in this paper. And we propose a modified grey Verhulst model. The new initial condition consists of the first item and the last item of a sequence generated from applying the first-order accumulative generation operator on the sequence of raw data. Weighted coefficients of the first item and the last item in the combination as the initial condition are derived from a method of minimizing error summation of square. We shows that the newly modified grey power model is an extension of the previous optimized GM(1,1) models and grey Verhulst models. The new optimized initial condition can express the principle of new information priority emphasized on in grey systems theory fully. The result of a numerical example indicates that the modified grey model presented in this paper can obtain a better prediction performance than that from the original grey model.

## 1. Introduction

Since 1982 (Deng, 1982), the grey systems theory has attracted a large number of researchers and has been utilized in a variety of fields such as natural science, social science and engineering science etc. Grey systems theory focuses mainly on such systems as partial information known and partial information unknown. With the rapid development of science and technology more and more systems have the same characteristic as grey systems. Solutions to these uncertain systems have become a great challenge for further development in associated fields. However, one of possible means can be provided by some approaches in grey systems theory.

GM(1,1) model is one of important models in grey models group. From the procedure of construction of GM(1,1) model we can find that the model is neither a differential equation nor a difference equation. In fact, it is an approximate model which has the characteristic both differential equation and difference equation. It is inevitable for the approximate model to result in some errors in practical applications. To increase prediction accuracy using GM(1,1) model a large number of researchers concentrate upon improvements of GM(1,1) model mainly in three aspects. On the one hand, a number of researchers focus mainly on improvements of grey derivative. Wang proposes a GM(1,1) direct modeling method with a step by step optimizing grey derivative's whitened values to unequal time interval sequence modeling. He has also proves that the new method still has the same characteristic of linear transformation consistency as the old method (Wang,

This paper was supported by Nanjing University of Aeronautics and Astronautics

Chen, Gao, & Chen, 2004). Mu presents a method to optimize the whitened values of grey derivative and constructs an unbiased GM(1,1) model. He also proves that the new model has the characteristic of law of whitened exponent (Mu, 2003a, 2003b).

On the other hand, a number of researchers focus mainly on improvements of background value. Tan presents a new method to construct the formulation of background value. The new model with the application of a new formulation of background value can simulate an exponential sequence well (Tan, 2000). Mu points out the reasons that the GM(1,1) model can not predict an exponential function perfectly through the analysis of mechanism of constructing the GM(1,1) model. And he proposes a new method to construct GM(1,1) model based on the improvement of background value (Mu, 2002). Mao utilizes a modified GM(1,1) model based on background value to predict vehicle fatality risk and obtain a better prediction performance (Mao & Chirwa, 2006).

Finally, improvements of the initial condition in the time response function are focused on by some researchers gradually. Dang proposes a method to improve grey models using the nth item of $X^{(1)}$ obtained by the first-order accumulative generation operator on a sequence of raw data as the starting condition of the grey differential model to increase prediction precision (Dang, Liu, & Chen, 2004). To make full use of new pieces of information in raw data and also reserve the initial condition in time response function of the original GM(1,1) model, Wang, Tang & Cao (2012) set a newly initial condition in time response function equaling to the arithmetic mean of the first item and the last item of $X^{(1)}$. Wang, Dang ,Li & Liu (2010) proposed an approach to optimize the initial condition in the time response function for original GM(1,1) model and derived the optimal weights of the first item and the last item of $X^{(1)}$ by the least square method. Xie proposes discretely grey prediction models and corresponding parameter optimization methods. And he also illustrates three classes of grey prediction models such as the starting-point fixed discrete grey model, the middle-point fixed discrete grey model and the ending-point fixed discrete grey model (Xie & Liu, 2008). Liu also presents a method to improve the prediction precision by optimization of the coefficient of exponential function (Liu & Lin, 2006). In addition to these three classes of improvements on GM(1,1) model, Hsu presents a grey model improved by the Bayesian analysis to predict output of integrated circuit industry (Hsu & Wang, 2007). Li also presents a method of combining GM(1,1) model and markov chain to predict the number of Chinese international airlines (Li, Yamaguchi, & Nagai, 2007).

These types of improvements for grey models may increase prediction precision in some practical applications. However, there still exists some space to improve prediction precision of GM(1,1) model. We present a novel approach to optimize the initial condition in time response function for general grey prediction model involving the GM(1,1) and grey Verhulst models as its special cases. The new optimization approach of the initial condition is comprised of the initial item in the time response function from the original GM(1,1) model (Deng, 2002; Liu & Lin, 2006) and the nth item of $X^{(1)}$ (Dang et al., 2004). And the corresponding weighted coefficients of the two parts forming the initial condition in new optimization method are derived from minimizing error summation of square in terms of the time response function. The combination of newly initial condition can express the principle of new information priority emphasized on in grey systems theory fully. Hence, we can make full use of new pieces of information in raw data to improve prediction precision with application of the new initial condition. Meanwhile, the new optimized model is also an extension of the original GM(1,1) model, the previous modified GM(1,1) models and grey Verhulst model. And the coefficients of weight in the optimized initial condition can change in the interval of zero and one. Especially, the newly optimized model could also be degenerated into the improved model presented by professor Dang (Dang et al., 2004), the original GM(1,1) model when the coefficients take zero or one respectively and the improved GM(1,1) models based on the optimization of initial condition ([1], [2]). Then we believe that the newly optimized model can be helpful to improve the prediction precision of grey prediction models including GM(1,1) model and Verhulst model.

The remaining parts of this paper are organized as following. A brief introduction to the original GM(1,1) model, the optimization method of the initial condition and a modified grey Verhulst model are given in Section 2. Section 3 will illustrate the application of new optimized method through a numerical example. Conclusions and the work focused mainly on in the future will be given in Section 4.

## 2. Optimization method of initial condition and mean values

### 2.1. The original GM(1,1) model

Let

$$X^{(0)} = \{x^{(0)}(1), x^{(0)}(2),..., x^{(0)}(N)\}$$

be a non-negative sequence of raw data, and its sequence of the first-order accumulative generation operator on $X^{(0)}$ is denoted as following (Deng, 2002; Liu &Lin, 2006),

$$X^{(1)} = \{x^{(1)}(1), x^{(1)}(2),..., x^{(1)}(N)\},$$

where $x^{(1)}(k) = \sum_{i=1}^{k} x^{(0)}(i), k = 1,..., N$.

From the sequence $X^{(1)}$ obtained from applying the first-order accumulative generation operator on sequence $X^{(0)}$, we can derive the sequence of generated mean value of consecutive neighbors in the following,

$$Z^{(1)} = \{z^{(1)}(2),\cdots, z^{(1)}(N)\};$$

where

$$z^{(1)}(k) = \frac{1}{2}(x^{(1)}(k) + x^{(1)}(k-1)), \ k = 2,..., N.$$

**Definition 1**(Deng, 2002; Liu & Lin, 2006): For a non-negative sequence of raw data $X^{(0)}$; $X^{(1)}$ is a newly generated sequence with the application of the first-order accumulative generation operator on $X^{(0)}$; $Z^{(1)}$ is a new sequence with the application of the generated mean value of consecutive neighbors operator on $X^{(1)}$, then the following

equation
$$x^{(0)}(k) + az^{(1)}(k) = b, \ k = 2,...,N$$

is a grey differential equation, also called GM(1,1) model. And the equation

$$\frac{dx^{(1)}}{dt} + ax^{(1)} = b$$

is the whitened equation of GM(1,1) model. If we let

$$Y = \begin{bmatrix} x^{(0)}(2) \\ x^{(0)}(3) \\ \vdots \\ x^{(0)}(n) \end{bmatrix}, \quad B = \begin{bmatrix} -z^{(1)}(2) & 1 \\ -z^{(1)}(3) & 1 \\ \vdots \\ -z^{(1)}(n) & 1 \end{bmatrix},$$

then the least squares estimate of the GM(1,1) model parameters[ ] is

$$[a,b]^T = (B^T B)^{-1} B^T Y.$$

The time response function of the whitened equation yields

$$x^{(1)}(t) = \left(x^{(0)}(1) - \frac{b}{a}\right)e^{-a(t-1)} + \frac{b}{a}, \ t \geq 1$$

And the time response equation of GM(1,1) is the following equation

$$x^{(1)}(k+1) = \left(x^{(0)}(1) - \frac{b}{a}\right)e^{-ak} + \frac{b}{a}, \ k = 1,...,N$$

The restored values of raw data are given below.

$$\hat{x}^{(0)}(k+1) = \hat{x}^{(1)}(k+1) - \hat{x}^{(1)}(k) = (1-e^a)\left(x^{(0)}(1) - \frac{b}{a}\right)e^{-ak}, \ k = 1,...,N$$

Professor Deng (2002) points out that the whitened equation and its corresponding time response function can not be derived from the GM(1,1) model directly, which are some approximate replacement. In fact, the time response function is a solution of the whitened function when the initial condition is $x^{(0)}(1)$. These approximate processing procedures also bring hints for us to improve prediction precision of GM(1,1) model. As the principle of new information priority is emphasized on in grey systems theory to improve prediction performance of data of small sample sizes. So how to improve performance in utilizing new information and keep consistent with the original GM(1,1) model are main contents in Section 2.2

## 2.2. Optimization of initial condition

Consider the following general grey prediction model:

$$x^{(0)}(k) + az^{(1)}(k) = b\left(z^{(1)}(k)\right)^{\alpha}, \; k = 2,...,N, \tag{1}$$

where

$$z^{(1)}(k) = \lambda x^{(1)}(k) + (1-\lambda)x^{(1)}(k-1), \; k = 2,...,N \tag{2}$$

and $\alpha \neq 1$, $\lambda \in [0,1]$ are parameters. The parameter $\lambda$ is called a mean value parameter. The model (1) is also called the GM(1,1) power model [7]. Often, $\lambda = 0.5$ is selected. The whitened equation of the above equation is

$$\frac{dx^{(1)}}{dt} + ax^{(1)} = b\left(x^{(1)}\right)^{\alpha} \tag{3}$$

and its general solution is as following ( Theorem 4.17 of [7]).

$$x^{(1)}(t) = \left\{e^{-(1-\alpha)at}\left[(1-\alpha)\int be^{(1-\alpha)at}dt + c\right]\right\}^{\frac{1}{1-\alpha}} = \left\{ce^{-(1-\alpha)at} + \frac{b}{a}\right\}^{\frac{1}{1-\alpha}}, \tag{4}$$

where $c$ is a constant, $a$ and $b$ are parameters to be derived from some estimation method. If $\alpha = 0$, then we have general form of GM(1,1) model

$$x^{(1)}(t) = ce^{-at} + \frac{b}{a} \tag{5}$$

from (4).
If $\alpha = 2$, then we have general form of Verhulst GM(1,1) model

$$x^{(1)}(t) = \left(ce^{at} + \frac{b}{a}\right)^{-1} \tag{6}$$

from (4).
For the general solution (4) if we let $t = 1$ and $t = n$, respectively, then we can obtain the following Eqs. (7) and (8).

$$x^{(1)}(1) = \left(ce^{-(1-\alpha)a} + \frac{b}{a}\right)^{\frac{1}{1-\alpha}}, \tag{7}$$

$$x^{(1)}(n) = \left(ce^{-(1-\alpha)an} + \frac{b}{a}\right)^{\frac{1}{1-\alpha}}. \tag{8}$$

From (7) and (8), we have

$$[x^{(1)}(1)]^{(1-\alpha)} = \left(ce^{-(1-\alpha)a} + \frac{b}{a}\right), \quad [x^{(1)}(n)]^{(1-\alpha)} = \left(ce^{-(1-\alpha)an} + \frac{b}{a}\right)$$

and then

$$\beta[x^{(1)}(1)]^{(1-\alpha)} + (1-\beta)[x^{(1)}(n)]^{(1-\alpha)} = \beta\left(ce^{-(1-\alpha)a} + \frac{b}{a}\right) + (1-\beta)\left(ce^{-(1-\alpha)an} + \frac{b}{a}\right)$$

$$= \beta ce^{-(1-\alpha)a} + (1-\beta)ce^{-(1-\alpha)an} + \frac{b}{a} = c\left[\beta e^{-(1-\alpha)a} + (1-\beta)e^{-(1-\alpha)an}\right] + \frac{b}{a},$$

where $\beta \in [0,1]$ is called the initial condition parameter. From the above, we obtain the constant determined by the initial condition

$$c = \left[\beta e^{-(1-\alpha)a} + (1-\beta)e^{-(1-\alpha)an}\right]^{-1} \left\{\beta[x^{(1)}(1)]^{(1-\alpha)} + (1-\beta)[x^{(1)}(n)]^{(1-\alpha)} - \frac{b}{a}\right\}. \tag{9}$$

For $\alpha = 0$, i.e. GM(1,1) model, we have

$$c = \left[\beta e^{-a} + (1-\beta)e^{-an}\right]^{-1} \left\{\beta[x^{(1)}(1)] + (1-\beta)[x^{(1)}(n)] - \frac{b}{a}\right\}, \tag{10}$$

which is the same as $c$ of [1]. When $\beta = 0.5$, $c$ defined by (10) is the same as $c$ of [2]. For $\alpha = 2$, i.e. grey Verhulst model, we have

$$c = \left[\beta e^{a} + (1-\beta)e^{an}\right]^{-1} \left\{\beta[x^{(1)}(1)]^{-1} + (1-\beta)[x^{(1)}(n)]^{-1} - \frac{b}{a}\right\}. \tag{11}$$

Let

$$f_1(c) = \sum_{k=1}^{n} \left|\hat{x}^{(1)}(k) - x^{(1)}(k)\right| = \sum_{k=1}^{n} \left|\left(ce^{-(1-\alpha)ak} + \frac{b}{a}\right)^{\frac{1}{1-\alpha}} - x^{(1)}(k)\right| \tag{12}$$

and

$$f_2(c) = \sum_{k=1}^{n} \left(\hat{x}^{(1)}(k) - x^{(1)}(k)\right)^2 = \sum_{k=1}^{n} \left(\left(ce^{-(1-\alpha)ak} + \frac{b}{a}\right)^{\frac{1}{1-\alpha}} - x^{(1)}(k)\right)^2. \tag{13}$$

We can obtain the estimate of $c$ so that it minimize the $f_1(c)$ or $f_2(c)$.

For $\alpha = 0$, i.e. GM(1,1) model, $\dfrac{\partial f_2(c)}{\partial c} = 0$ if and only if $\sum\limits_{k=1}^{n}\left(ce^{-ak} + \dfrac{b}{a} - x^{(1)}(k)\right)e^{-ak} = 0$,

from which we obtain the least square estimate

$$c = -\left(\sum_{k=1}^{n} e^{-2ak}\right)^{-1} \sum_{k=1}^{n} \left(\frac{b}{a} - x^{(1)}(k)\right) e^{-ak} . \tag{14}$$

From (10) and (14), we obtain the initial condition parameter $\beta$ as following.

$$\beta = \frac{\left(x^{(1)}(n) - b/a\right) \sum_{k=1}^{n} e^{-2ak} - e^{-an} \sum_{k=1}^{n} \left(x^{(1)}(k) - b/a\right) e^{-ak}}{\left(e^{-a} - e^{-an}\right) \sum_{k=1}^{n} \left(x^{(1)}(k) - b/a\right) e^{-ak} + \left(x^{(1)}(n) - x^{(1)}(1)\right) \sum_{k=1}^{n} e^{-2ak}}, \tag{15}$$

which is the same as one of [1]. Then we can obtain the following optimized time response function for GM(1,1) model from (5), (10) and (15):

$$\hat{x}^{(1)}(t) = \left[\beta x^{(1)}(1) + (1-\beta) x^{(1)}(n) - \frac{b}{a}\right] \left[\beta e^{-a} + (1-\beta) e^{-an}\right]^{-1} e^{-at} + \frac{b}{a} \tag{16}$$

The restored values for GM(1,1) model can be derived by first-order inverse accumulate generation operator on the time response function (16):

$$\hat{x}^{(0)}(t) = \hat{x}^{(1)}(t) - \hat{x}^{(1)}(t-1)$$
$$= \left(1 - e^{a}\right) \left[\beta x^{(1)}(1) + (1-\beta) x^{(1)}(n) - \frac{b}{a}\right] \left[\beta e^{-a} + (1-\beta) e^{-an}\right]^{-1} e^{-at} \tag{17}$$

The time response function for grey Verhulst model is obtained by (6) and (11) as follows.

$$\hat{x}^{(1)}(t) = \frac{a\left[\beta e^{a} + (1-\beta) e^{an}\right] x^{(1)}(1) x^{(1)}(n)}{\left[\beta x^{(1)}(n)a + (1-\beta) x^{(1)}(1)a - b x^{(1)}(1) x^{(1)}(n)\right] e^{at} + b\left[\beta e^{a} + (1-\beta) e^{an}\right] x^{(1)}(1) x^{(1)}(n)}, \tag{18}$$

where parameter $\beta$ can be derived by substituting $c$ minimizing $f_1(c)$ or $f_2(c)$ from (11). If $\beta = 1$, then we have

$$\hat{x}^{(1)}(t) = \frac{a x^{(1)}(1)}{b x^{(1)}(1) + \left(a - b x^{(1)}(1)\right) e^{a(t-1)}} , \tag{19}$$

which is just the time response function of the original grey Verhulst model [7].
The restored values for grey Verhulst model can be derived by first-order inverse accumulate generation operator from the time response function (19).

### 2.3. Modified grey Verhulst Model

From (1) and (2), the following grey Verhulst model is obtained.

$$x^{(0)}(k) + a[x^{(1)}(k-1) + \lambda x^{(0)}(k)] = b[x^{(1)}(k-1) + \lambda x^{(0)}(k)]^2, \ k = 2,...,N, \quad (20)$$

where $\lambda \in [0,1]$ is the mean value parameter.
Solving the quadratic equation (20) with $x^{(0)}(k)$, we have

$$x^{(0)}(k) = \frac{1 + a\lambda - 2b\lambda x^{(1)}(k-1)}{2b\lambda^2} - \frac{\sqrt{[2b\lambda x^{(1)}(k-1) - a\lambda - 1]^2 + 4b\lambda^2 x^{(1)}(k-1)[a - bx^{(1)}(k-1)]}}{2b\lambda^2} \quad (21)$$

The model (21) is called the modified grey Verhulst model. We can obtain the restored values of grey Verhulst model from (21) by substituting (18) instead of $x^{(1)}(k)$.

## 2.4. Optimization of mean values and model parameters

In section 2.2, we considered the optimization of initial condition for general form of grey prediction model and derived some optimal initial condition parameters. In this section is considered the optimization of mean values and model parameters for the general grey prediction model defined by (1) and (2).
Let

$$f_p(a,b,\lambda) = \|x^0 - x(a,b,\lambda)\|_p \quad (22)$$

and

$$\hat{f}_p(a,b,\lambda) = \left\| \left( \frac{x^{(0)}(2) + az^{(1)}(2) - b(z^{(1)}(2))^\alpha}{x^{(0)}(2)}, \cdots, \frac{x^{(0)}(N) + az^{(1)}(N) - b(z^{(1)}(N))^\alpha}{x^{(0)}(N)} \right) \right\|_p, \quad (23)$$

where
$$z^{(1)}(k) = \lambda x^{(1)}(k) + (1-\lambda)x^{(1)}(k-1), \ k = 2,...,N,$$
$$x^0 = (x^{(0)}(2),...,x^{(0)}(N))^T,$$
$$x(a,b,\lambda) = -az(\lambda) + bz^\alpha(\lambda), \quad (24)$$
$$z(\lambda) = (z^{(1)}(2),...,z^{(1)}(N))^T, \ z^\alpha(\lambda) = ((z^{(1)}(2))^\alpha,...,(z^{(1)}(N))^\alpha)^T$$

and $\|\circ\|_p$ is $p$-norm of vector ($p = 1, 2$ or $\infty$).
The parameters $a$, $b$ and $\lambda$ of (1) and (2) is determined so that residual function $f_p(a,b,\lambda)$ or $\hat{f}_p(a,b,\lambda)$ is minimized. The minimization problem of $f_p(a,b,\lambda)$ or $\hat{f}_p(a,b,\lambda)$ is non-convex programming problem in general and is difficult to find global optimal solution. Therefore, we can obtain the parameters $a$, $b$ and $\lambda$ in such way that $f_p(a,b,\lambda)$ or $\hat{f}_p(a,b,\lambda)$ is minimized with $a$ and $b$ for fixed $\lambda$.
Consider the case of $\alpha = 0$, i.e. GM(1,1) model. Then, from (24), we have

$$x(a,b,\lambda) = -az(\lambda) + be, \quad e = (1,1,...,1)^T.$$

Let $y = (a,u)^T$, $f_p^\lambda(a,u) \equiv f_p(a,u,\lambda)$ for fixed $\lambda$. Suppose that we make use of the least square method, i.e., $f_p(a,u,\lambda) = \|x^0 - x(a,u,\lambda)\|_2^2$ instead of $f_p(a,u,\lambda) = \|x^0 - x(a,u,\lambda)\|_p$. In this case, $y = (a,b)^T$ for fixed $\lambda$ is obtained as follows.

$\dfrac{\partial f_2^\lambda}{\partial a} = 0, \dfrac{\partial f_2^\lambda}{\partial b} = 0$ if and only if

$$A(\lambda)y = b(\lambda), \tag{25}$$

where $A(\lambda) = \begin{pmatrix} \|z(\lambda)\|^2 & -e^T z(\lambda) \\ -e^T z(\lambda) & \|e\|^2 \end{pmatrix}$ and $b(\lambda) = \begin{pmatrix} -z(\lambda)^T x^0 \\ e^T x^0 \end{pmatrix}$.

When $\lambda = 0.5$, the solution of (25) is just the least square estimate for the original GM(1,1) model, i.e. $A(0.5) = B^T B$, $b(0.5) = B^T Y$ and $y = (B^T B)^{-1} B^T Y$.

Since $z(\lambda)$ and $e$ is linearly independent, i.e., $z(\lambda) \neq \gamma e$ for every non-zero $\gamma$, $A(\lambda)$ is non-singular and the least square estimate is obtained by $y = A(\lambda)^{-1} b(\lambda)$. Since

$$A(\lambda)^{-1} = \dfrac{1}{\det(A(\lambda))} \begin{pmatrix} \|e\|^2 & e^T z(\lambda) \\ e^T z(\lambda) & \|z(\lambda)\|^2 \end{pmatrix}, \tag{26}$$

we have

$$\begin{pmatrix} a(\lambda) \\ u(\lambda) \end{pmatrix} = y(\lambda) = A(\lambda)^{-1} b(\lambda) = \dfrac{1}{\det(A(\lambda))} \begin{pmatrix} -\|e\|^2 z(\lambda)^T x^0 + (e^T z(\lambda))(e^T x^0) \\ -(z(\lambda)^T x^0)(e^T z(\lambda)) + \|z(\lambda)\|^2 e^T x^0 \end{pmatrix}. \tag{27}$$

## 3. Numerical example

We consider two examples to demonstrate the advantage of the optimized grey power model. The prediction performance is evaluated according to the relative error defined by

$$\dfrac{x^{(0)}(k) - \hat{x}^{(0)}(k)}{x^{(0)}(k)} \times 100 \ (\%), \ k = 2, \cdots, N$$

**Example 1** [1]. Let's consider the following data sequence generated by $f(t) = 2e^{0.4t}$, $t=1,\ldots,7$.

{2.9836, 4.4511, 6.6402, 9.9061, 14.7781, 22.0464, 32.8893}.

For the construction of grey model, we take the first five data of the sequence as $X^{(0)}$, i.e.

$X^{(0)} = \{2.9836, 4.4511, 6.6402, 9.9061, 14.7781\}$.

The results of numerical simulation by different grey models are shown in Table 1.

Table 1. The comparison of different GM(1,1) models

| Original data [1] | | GM(1,1) | | GM(1,1) (optimal $\beta$, error 2-norm) | | OptimalGM(1,1) (optimal $\beta$, error 1-norm) | | OptimalGM(1,1) (optimal $\beta$, error 2-norm) | | OptimalGMpower ($\alpha$= -0.1, optimal $\beta$, error 2-norm) | |
|---|---|---|---|---|---|---|---|---|---|---|---|
| t | Real value | Predicted value | relative error (%) | Predicted value | relative error (%) | Predicted value | relative error (%) | Predicted value | relative error (%) | Predicted value | relative error (%) |
| 2 | 4.4511 | 4.3804 | 1.5884 | 4.4562 | -0.1152 | 4.4514 | -0.0064 | 4.4480 | 0.0695 | 4.5241 | -1.6407 |
| 3 | 6.6402 | 6.5006 | 2.1027 | 6.6131 | 0.4080 | 6.6586 | -0.2775 | 6.6536 | -0.2014 | 6.6151 | 0.3777 |
| 4 | 9.9061 | 9.6469 | 2.6162 | 9.8139 | 0.9304 | 9.9603 | -0.5476 | 9.9528 | -0.4713 | 9.8328 | 0.7702 |
| 5 | 14.7781 | 14.3162 | 3.1257 | 14.5640 | 1.4487 | 14.8992 | -0.8198 | 14.8879 | -0.7432 | 14.7167 | 0.4153 |
| 6 | 22.0464 | 21.2454 | 3.6332 | 21.6132 | 1.9650 | 22.2871 | -1.0919 | 22.2702 | -1.0151 | 22.1046 | -0.2638 |
| 7 | 32.8893 | 31.5285 | 4.1376 | 32.0743 | 2.4781 | 33.3383 | -1.3653 | 33.3130 | -1.2883 | 33.2672 | -1.1491 |
| mean | | | 2.45768 | | 1.01644 | | -0.5869 | | -0.5214 | | -0.2172 |
| r | | 0.5 | | 0.5 | | 0.4499999 | | 0.4499999 | | 0.483343 | |
| $\beta$ | | 1 | | 0.522684 | | 0.624079 | | 0.521557 | | 0.231639 | |

In Table 1, 2 and 3, the optimal GM(1,1) model is the GM(1,1) model based on the minimization of the error function (23) with p=1, i.e. the absolute relative error summation. As seen from the table, the modified GM(1,1) model based on (15), (16), in which optimal $\beta$ is obtained by minimization of $f_2(c)$, is better than the original GM(1,1) model and the optimal GM(1,1) models are much better than the modified GM(1,1) model based on the optimal $\beta$. Especially, the optimal grey power model with $\alpha$ = -0.1 showed the best performance of prediction.

**Example 2.** Assume that 1992~2004 oil production amount ($10^4$ t) per year is given in the following sequence.

{5565.83, 5590.19, 5600.52, 5600.69, 5600.87, 5600.92, 5570.38, 5450.19, 5300.09, 5150.16, 5013.10, 4840.03, 4640.03}

We forecasted the 2001~2004 annual oil production amounts based on 1992~2000 data

$X^{(0)} = \{5565.83, 5590.19, 5600.52, 5600.69, 5600.87, 5600.92, 5570.38, 5450.19, 5300.09\}$

The results of numerical simulation by different grey models are shown in Table 2.

Table 2. The comparison of different GM(1,1) models

| Original data | | GM(1,1) | | GM(1,1) (optimal β, error 2-norm) | | OptimalGM(1,1) (optimal β, error 2-norm) | | OptimalGM(1,1) (optimal β, error 1-norm) | | OptimalGM(1,1) (optimal β, r-error 1-norm) | |
|---|---|---|---|---|---|---|---|---|---|---|---|
| year | Real value | Predicted value | relative error (%) | Predicted value | relative error (%) | Predicted value | relative error (%) | Predicted value | relative error (%) | Predicted value | relative error (%) |
| 1992 | 5565.83 | 5565.83 | 0 | 5565.83 | 0 | 5565.83 | 0 | 5565.83 | 0 | 5565.83 | 0 |
| 1993 | 5590.19 | 5658.52 | -1.2222 | 5658.51 | -1.2221 | 5641.12 | -0.9111 | 5641.34 | -0.9150 | 5641.54 | -0.9186 |
| 1994 | 5600.52 | 5624.01 | -0.4195 | 5624.01 | -0.4194 | 5606.76 | -0.1114 | 5606.98 | -0.1153 | 5607.17 | -0.1188 |
| 1995 | 5600.69 | 5589.72 | 0.1957 | 5589.72 | 0.1958 | 5572.60 | 0.5014 | 5572.82 | 0.4976 | 5573.01 | 0.4941 |
| 1996 | 5600.87 | 5555.64 | 0.8075 | 5555.64 | 0.8076 | 5538.65 | 1.1108 | 5538.87 | 1.1069 | 5539.06 | 1.1034 |
| 1997 | 5600.92 | 5521.77 | 1.4132 | 5521.76 | 1.4132 | 5504.91 | 1.7141 | 5505.13 | 1.7102 | 5505.32 | 1.7068 |
| 1998 | 5570.38 | 5488.10 | 1.4770 | 5488.09 | 1.4771 | 5471.38 | 1.7772 | 5471.59 | 1.7734 | 5471.78 | 1.7700 |
| 1999 | 5450.19 | 5454.64 | -0.0816 | 5454.63 | -0.0815 | 5438.05 | 0.2227 | 5438.26 | 0.2188 | 5438.45 | 0.2154 |
| 2000 | 5300.09 | 5421.38 | -2.2888 | 5421.38 | -2.2888 | 5404.92 | -1.9778 | 5405.13 | -1.9818 | 5405.32 | -1.9854 |
| mean | | | -0.0132 | | -0.0131 | | 0.2584 | | 0.2549 | | 0.2518 |
| 2001 | 5150.16 | 5388.33 | -4.6245 | 5388.32 | -4.6243 | 5371.99 | -4.3073 | 5372.20 | -4.3113 | 5372.39 | -4.3150 |
| 2002 | 5013.10 | 5355.47 | -6.8295 | 5355.46 | -6.8295 | 5339.27 | -6.5064 | 5339.47 | -6.5104 | 5339.66 | -6.5141 |
| 2003 | 4840.03 | 5322.82 | -9.9749 | 5322.81 | -9.9748 | 5306.74 | -9.6427 | 5306.95 | -9.6470 | 5307.13 | -9.6508 |
| 2004 | 4640.03 | 5290.36 | -14.016 | 5290.36 | -14.016 | 5274.41 | -13.671 | 5274.62 | -13.676 | 5274.80 | -13.680 |
| mean | | | -8.8612 | | -8.8611 | | -8.5321 | | -8.5363 | | -8.5401 |
| r | | 0.5 | | 0.5 | | 0 | | 0 | | 0 | |
| β | | 1 | | 0.499969 | | 0.494846 | | 0.758834 | | 1 | |

As seen from the table, the modified GM(1,1) model based on (15) and (16) has almost same performance with the original GM(1,1) model, while the optimal GM(1,1) models are much better than the modified GM(1,1) model based on the optimal β. The optimal GM(1,1) model based on minimization of $\hat{f}_1(a,b,\lambda)$ and $f_2(c)$ showed the best performance of prediction.

## 4. Discussion

In this paper we propose an optimized method of the initial condition in grey power model having GM(1,1) model and Verhulst model as its parts to improve its prediction accuracy. We have the initial condition of the grey power model consisted of the first item $x^{(1)}(1)$ and the last item $x^{(1)}(n)$ of a new sequence $X^{(1)}$ generated from applying the first-order accumulative generation operator on the sequence of raw data $X^{(0)}$. From minimizing the squared error summation or absolute error summation, we can derive the weight coefficients $β$ and $1-β$ for $x^{(1)}(1)$ and $x^{(1)}(n)$ in the initial condition. The result of numerical examples indicate that the modified model can improve prediction accuracy of GM(1,1) model significantly. The numerical examples show that the optimal estimates $a$, $b$ and $λ$ by minimization of the absolute relative error summation can improve much the prediction accuracy of the grey power model. We can also show that the modified grey power model is an extension of the original GM(1,1) model proposed by professor Deng (Deng, 1982, 1989, 2002), professor Liu (Liu, Dang, & Fang, 2004; Liu & Lin, 2006), the modified GM(1,1) model presented by professor Dang et al. (2004), the recent optimized GM(1,1) model of Wang, Dang ,Li & Liu (2010) and the grey Verhulst model according to values of $α$ and $β$ . The modified grey model can also express the principle of new information priority emphasized on in grey systems theory fully. Then adequate information can be extracted from the sequence of small sample sizes when we apply the modified grey prediction model.

Some modifications such as optimization of the grey derivative, optimization of the power index $α$ and combinations of distinct methods will be focused mainly on to improve prediction accuracy of the general grey prediction model in future work.